\documentclass[letterpaper,12pt]{article} 
\input{epsf}
\usepackage{amsmath,amsfonts} 

 \normalsize

\def\Pcm#1{{\mathcal{#1}}}
\def\nn{\nonumber}
\def\er#1{eqn.~\eqref{#1}}
\newcommand{\re}{\Re e \ }

\newcommand{\Dp}{\textup{D}p}
\newcommand{\td}{\textup{d}}
 
\title{Dp-brane Tension from Tachyons and B-field in Vacuum String Field Theory}
\author{P.~Matlock$^a$, R.C.~Rashkov$^b$, K.S.~Viswanathan$^a$ and Y.~Yang$^a$ \cr\cr
     ${}^a$ \small \emph{Department of Physics, Simon Fraser University,}\cr
            \small \emph{Burnaby BC, Canada}\cr 
            \small \texttt{pwm@sfu.ca},\quad \texttt{kviswana@sfu.ca},
            \quad \texttt{yyangc@sfu.ca}\cr 
     ${}^b$ \small \emph{Department of Physics, Sofia University}\cr
            \small \emph{1164 Sofia, Bulgaria}\cr
            \small \texttt{rash@phys.uni-sofia.bg} }
\date{March 7$^{\textup{th}}$, 2002}
\begin{document} 
\maketitle 
\begin{abstract}
  We consider tachyonic string-field fluctuations about a 
D$p$-brane background in the geometrical (CFT) formulation 
of Vacuum String Field Theory. We then extend this analysis
to the case of a background $B$-field. We find that the standard
results for D-brane tension are reproduced in both cases.
\end{abstract}
\newpage
\section{Introduction}
 Vacuum String Field Theory~\cite{RSZ5} is believed to represent Witten's cubic open 
string field theory~\cite{Witten1} (see also \cite{GJ1,GJ2,KP,GT1,GT2}) near the tachyon vacuum.
Sen conjectured that unstable D-brane configurations may collapse into other
stable configurations via condensation of the tachyon~\cite{Sen,Sen2}. In particular, the
D25-brane background of open bosonic string theory is expected to condense into a
non-perturbative tachyon vacuum~\cite{RZ,RSZ1}.
In VSFT the usual BRST operator $Q$, whose cohomology classes represent physical 
states in the
full theory, is replaced by a new operator $\Pcm{Q}$ which operates only on ghost degrees of freedom
and has trivial cohomology. Therefore there are no perturbative physical states in 
VSFT, and moreover due to this simplification it is possible to find classical solutions 
to the string-field equation of motion~\cite{KP,GT1,RSZ2,RSZ5}. 
In Witten's original notation~\cite{Witten1}, the action is given by
\[S[\Psi]=-\frac12\int\left(\Psi\star Q \Psi +g\frac23 \Psi\star\Psi\star\Psi \right),\]
so that the equation of motion for the string-field $\Psi$ is $Q\Psi=\Psi \star \Psi$.

 The problem of tachyon fluctuations and the resulting D25-brane tension in the context of VSFT 
was first considered by Hata and Kawano~\cite{HK}, and by Rastelli, Sen and Zwiebach~\cite{RSZ6}.
In the former paper the D25-brane tension did not seem to be compatible with the standard result,
and in the latter doubt was expressed as to the validity of the linearised equations 
of motion for the tachyon perturbation, when the BPZ-product with the perturbation state (which is
slightly outside the Fock-space) is taken. The analysis by two of the present authors in \cite{RV}
clarified the limiting procedure involved when taking star- and BPZ-products, and showed 
that the linearised equation of motion is indeed valid in the `strong' sense, obtaining a result
for the D25-brane tension compatible with the assertion that the sliver solution represents 
a single D25-brane.

Here we consider tachyon fluctuations about a D$p$-brane solution for arbitrary $p$ 
and in this way evaluate the tension of the D-brane. We make use of boundary condition-changing
twist operators, as suggested in \cite{RSZ4}, in the `geometrical' conformal field theory 
approach. We find the correct ratios of tensions between branes of differing dimension.
Our results may be compared to those found in the recent paper \cite{Okuyama}
 which were obtained using the algebraic (oscillator) \cite{Muk,GJ1,GJ2} approach.
We also investigate the effect of a $B$-field and again obtain the expected ratios of tensions.

This paper is organised as follows. In section \ref{Dpbrane} we discuss the construction of
perturbative states around a D-brane solution, in particular focussing our attention on 
the tachyon field. Next, in section \ref{VSFTP} we construct this perturbative tachyon state 
around a D$p$-brane configuration and evaluate the resulting tension. 
 Section \ref{Bfield} contains an analysis
of the effect of a $B$-field background, and we again obtain the standard expression for the
tension of a non-commutative D-brane. We conclude with some comments in section \ref{discuss}.

\section{D$p$-brane}
\label{Dpbrane}

In Fock-space notation, as used in the geometrical approach of Rastelli, Sen and 
Zwiebach~\cite{RSZ5}, the action is given by
\begin{equation}
\label{VSFTact}
S[\Psi]=-\kappa\left[\frac12\big\langle\Psi\big|\Pcm{Q}\Psi\big\rangle+\frac13\big\langle\Psi\big|\Psi * \Psi\big\rangle \right]
,\end{equation}
and the equation of motion is written
\begin{equation}
\label{EOM}
\Pcm{Q}\left|\Psi\right\rangle = \left|\Psi * \Psi \right\rangle
.\end{equation}
In VSFT the operator $\Pcm{Q}$ affects only the ghost sector 
of the theory; the equation of motion thus 
factorises and the matter part becomes
\begin{equation}
\left|\Psi_m\right\rangle = \left|\Psi_m * \Psi_m \right\rangle
.\end{equation}
It is conjectured that the ghost part of the solution is universal~\cite{Sen}.

One solution $\Xi$ to the matter equation of motion, called the sliver state, is believed to 
correspond to a D25-brane. This state is defined by
\begin{equation}
\label{sliver}
\big\langle \Xi \big| \phi \big\rangle = \lim_{n \rightarrow \infty} 
                       \Pcm{N}\big\langle f \circ \phi(0) \big\rangle_{C_n}
\end{equation}
where on the RHS the brackets denote a correlation function in the matter BCFT
on the space $C_n$. $C_n$ is taken to be the semi-infinite cylinder 
obtained by identifying $\re z \equiv \re z + n\pi/2$ in the complex $z$ upper half plane.
$\phi$ is an operator representing an arbitrary state in the Fock space 
and $|\phi\rangle$ is the corresponding state obtained by inserting $\phi$
at $z=0$ in the `local patch' $-\pi/4 < \re z < \pi/4$. We refer the reader to 
\cite{RSZ5} for details of this construction.

In order to extend this to the case of a D$p$-brane, we need to introduce 
boundary condition-changing twist operators~\cite{RSZ4,Muk}. These are
$\sigma^+$ and $\sigma^-$ and are inserted at $z=\pm\pi/4\pm\epsilon$.
This effectively imposes Neumann boundary conditions on $-\pi/4 < z < \pi/4$ and
Dirichlet conditions on the rest of the boundary, $\pi/4 < z < (2n-1)\pi/4$.
The correlator in \er{sliver} is taken over the 26 independent conformal fields
$X^\mu$. In this case, we define
\begin{equation}
\label{dps}
\big\langle \Dp \big| \phi \big\rangle = \lim_{n \rightarrow \infty} 
                       \Pcm{N}\big\langle f \circ \phi(0) \big\rangle^{\sigma \perp}_{C_n}
,\end{equation}
where the superscript indicates the presence of the $\sigma$ operators in the 25$-p$ Dirichlet directions,
while the $p+1$ Neumann directions are unchanged.
The state $|\Dp\rangle$  so defined represents 
a D$p$-brane~\cite{RSZ4} and it satisfies the
equation of motion \eqref{EOM}, provided that it is renormalised as follows.
When we take the star-product of the sliver with itself, we will obtain a 
short-distance singularity from the proximity of $\sigma^+$ from one sliver, and
$\sigma^-$ from the other. As noted in~\cite{RSZ4}, the leading term of the operator 
expansion will have no operator content, and so this singularity will only 
contribute $(1/2\epsilon)^h$ to the product, where $h$ depends on the conformal
dimensions of the $\sigma$-operators. This divergent factor may simply be absorbed
into the definition of the sliver state.

\section{VSFT Perturbations and D$p$-brane Tension}
\label{VSFTP}
We wish to start with a background string-field solution $\Phi_0=\Phi_{\textup{ghost}}\otimes\Phi$ 
and consider perturbations parametrised by fields. We follow here the procedure used 
in \cite{RV}, generalised to a the case of a D$p$-brane.
 As in \cite{RSZ6}, we use the perturbative expansion
\begin{equation}
\big| \Psi \big\rangle =  \big| \Phi_g \big\rangle \otimes
        \left\{
          \big|\Phi \big\rangle + \big|T\big\rangle + \cdots
        \right\}
,\end{equation}
where $|T\rangle$ is a tachyon excitation and terms corresponding to vector 
and higher excitations follow.
The tachyon perturbation is
\begin{equation}
 \big|T\big\rangle = \int \td k n^{-k_\parallel^2} T(k) \big|\chi(k)\big\rangle 
,\end{equation}
with $T(k)$ the momentum-space tachyon field. We use $k_\parallel$ to refer to
directions longitudinal to the D-brane, and $k_\perp$ for transverse directions.

We may insert this expansion for $|\Psi\rangle$ into the action \eqref{VSFTact}
to obtain 
\begin{equation}
\label{Tact}
S[T]=S[\Phi_g\otimes\Phi]-\big\langle\Phi_g\big|\Pcm{Q}\Phi_g\big\rangle
                  \left[
                    \frac12 \big\langle T\big|T\big\rangle
                    - \big\langle\Phi\big|T * T\big\rangle
                    + \frac13 \big\langle T\big|T * T\big\rangle
                  \right]
\end{equation}
whence the linearised equation of motion for $|T\rangle$ can be obtained;
\begin{equation}
\label{leom}
 \big|\chi(k)\big\rangle = \big|\chi(k) * \Phi \big\rangle + \big|\Phi * \chi(k)\big\rangle
.\end{equation}
In this section we take as the background solution the D$p$-brane state $|\Phi\rangle=|\Dp\rangle$.
It was shown in \cite{RV} that the linearised equation of motion for a perturbation $\chi(k)$
about the sliver state holds even when
the BPZ-product with another solution $\chi(k')$ is taken, that is
\begin{equation}
\label{strongeom}
 \big\langle\chi(k)\big|\chi(k')\big\rangle = 
     \big\langle \chi(k) * \Xi + \chi(k) * \Xi \big| \chi(k') \big\rangle
.\end{equation}
Equation \eqref{leom} has been referred to as the `weak' equation of motion, with \er{strongeom}
being a stronger version, since $\Xi$ and thus $\chi$ are not quite Fock space states.

The linearised equation of motion \eqref{leom} for a tachyon perturbation $\chi(k)$
about the sliver state $\Xi$ was solved in \cite{HK} and this
solution was expressed in the CFT language in \cite{RSZ6}. In the present case of
a D$p$-brane, the solution may be written
\begin{equation}
\label{chi}
\big\langle \chi(k) \big| \phi \big\rangle = \lim_{n \rightarrow \infty} 
                       \Pcm{N}n^{2k_\parallel^2}\big\langle f \circ \phi(0) e^{ik\cdot X}(n\pi/4)\big\rangle^{\sigma \perp}_{C_n}
,\end{equation}
so that the momentum degrees of freedom are carried by a tachyon vertex operator inserted 
diametrically opposite the $\phi$-insertion puncture. The factor $n^{2k_\parallel^2}$ is inserted to 
compensate for a factor $n^{-2k_\parallel^2}$ which will come from the correlator.

Let us now show that the state in \er{chi} satisfies the equation of motion \eqref{strongeom}.
Computing first the LHS, we find
\begin{equation}
\label{Cappcc}
\big\langle\chi(k)\big|\chi(k')\big\rangle=\Pcm{N}^2\lim_{n,m\rightarrow\infty}
 n^{2k_\parallel^2}m^{2{k'}_\parallel^2}
 \left\langle e^{ik\cdot X}(0)e^{ik'\cdot X}\big((n+m-2)\frac\pi4\big)\right\rangle^{\sigma\perp}_{C_m+n-2}
.\end{equation}
This is calculated in the appendix and is given by \er{Achichi},
\begin{equation}
\label{chichi}
\big\langle\chi(k)\big|\chi(k')\big\rangle
 =\Pcm{K}
 \lim_{n\rightarrow\infty} n^{2k_\parallel^2}
 2^{2k_\parallel^2}
 (2\pi)^{26}\delta(k_\parallel+k_\parallel')e^{ix_0\cdot (k_\perp+k_\perp')}
.\end{equation}
Shifting our attention to the RHS of \er{strongeom}, we may express either of the two terms as
\begin{equation}
\label{xcc}
\big\langle \Xi * \chi(k) \big| \chi(k') \big\rangle = \Pcm{N}^2 \lim_{n_1,n_2,n_3\rightarrow\infty}
  n_2^{2k_\parallel^2}n_3^{2{k'}_\parallel^2}
 \left\langle e^{ik\cdot X}(0)e^{ik'\cdot X}\big((n_2+n_3-2)\frac\pi4\big)
 \right\rangle^{\sigma\perp}_{C_{n_1+n_2+n_3-3}}
.\end{equation}
Evaluating this correlator in the same way we find
\begin{equation}
\label{Xichichi}
\big\langle \Xi * \chi(k) \big| \chi(k') \big\rangle
 = 
 \Pcm{K}
 \lim_{n\rightarrow\infty} n^{2k_\parallel^2}
 2^{k_\parallel^2} 
(2\pi)^{26}\delta(k_\parallel+k_\parallel')e^{ix_0\cdot (k_\perp+k_\perp')}
.\end{equation}
Substituting \er{chichi} and \er{Xichichi} into the equation of motion \eqref{strongeom},
it reduces to
\begin{equation}
\label{kseo}
k_\parallel^2=1
.\end{equation}
We see that the tachyon living on the D$p$-brane satisfies the strong equation of 
motion, provided that it is on-shell. 
This is to be compared with the case of a D25-brane~\cite{RV}.
Off-shell we have
\begin{equation}
\label{OSseom}
 2^{1-k_\parallel^2}\big\langle\chi(k)\big|\chi(k')\big\rangle = 
     \big\langle \chi(k) * \Xi + \chi(k) * \Xi \big| \chi(k') \big\rangle
.\end{equation}

To calculate the D$p$-brane tension, we first examine the quadratic term $S^{(2)}$ of
the action \eqref{Tact} to fix the normalisation of the tachyon field $T(k)$.
Using \er{OSseom} we may write 
\begin{equation}
S^{(2)}=-\frac12 \big\langle\Psi_g\big|\Pcm{Q}\Psi_g\big\rangle
   \int \td k \td k' T(k) T(k') \left(1-2^{1-k_\parallel^2}\right)
\big\langle\chi(k)\big|\chi(k')\big\rangle
.\end{equation}
After substituting the product from \er{chichi}, we have
\begin{eqnarray}
S^{(2)}&=&-\frac12 \big\langle\Psi_g\big|\Pcm{Q}\Psi_g\big\rangle \Pcm{K} 
    \int \td^{p+1} k_\parallel \td^{p+1} k'_\parallel
    \left(1-2^{1-k_\parallel^2}\right)
    2^{k_\parallel^2} 
    (2\pi)^{26}
    \delta(k_\parallel+k_\parallel')
   \nn\\
   && \left\{
    \int \td^{25-p} k_\perp \td^{25-p} k'_\perp
    e^{ix_0\cdot(k_\perp+k_\perp')} 
    T(k) T(k')
    \right\}
\end{eqnarray}
Taking $T(k)$ near on-shell, $k_\parallel^2\approx1$, we may write the $k_\parallel$-dependent 
factor as $(k_\parallel^2-1)4\log 2$. We see that a re-definition of the tachyon field
\begin{equation}
\label{newtach}
  T_{\Dp}(k_\parallel) \equiv \sqrt{\Pcm{K}(2\pi)^{25-p}\big\langle\Psi_g\big|\Pcm{Q}\Psi_g\big\rangle 4\log2 } 
   \int \td^{25-p} k_\perp 
    e^{ix_0 \cdot k_\perp} T(k)
\end{equation}
would cast $S^{(2)}$ into the canonical form
\begin{equation}
S^{(2)}=-\frac12 (2\pi)^{p+1} \int \td^{p+1} k_\parallel ({k_\parallel}^2-1) T_{\Dp}
   ( k_\parallel )T_{\Dp}(-k_\parallel ) 
.\end{equation}

The cubic term in the action involves
\begin{eqnarray}
\label{chichichi}
 && \big\langle\chi(k_1)\big|\chi(k_2) * \chi(k_3)\big\rangle \nn\\
 && = \left\langle e^{ik_1\cdot X}(0)
                e^{ik_2\cdot X}\big((n_1+n_2-2)\frac\pi4\big)
                e^{ik_3\cdot X}\big(-(n_1+n_3-2)\frac\pi4\big)
\right\rangle^{\sigma\perp}_{C_{n_1+n_2+n_3-3}}
.\end{eqnarray}
Once again mapping to the unit disc, we may use Wick's theorem and the tachyon
correlator from \er{DiscTwoCorr} to find
\begin{equation}
\big\langle\chi(k_1)\big|\chi(k_2) * \chi(k_3)\big\rangle = \lim_{n\rightarrow \infty}
   2n^{\sum k_\parallel^2} 
   (2\pi)^{26}
   e^{ix_0\cdot(\sum k_\perp)} 
   \delta\left(\sum k_\parallel\right) 
\end{equation}
Substituting this result into the action, the cubic term $S^{(3)}$ is
\begin{eqnarray}
S^{(3)}&=&-\frac23 \big\langle\Psi_g\big|\Pcm{Q}\Psi_g\big\rangle \Pcm{K} 
    \int \td k_{1,2,3\parallel}
    (2\pi)^{26}
    \delta\left(\sum k_\parallel\right)
   \nn\\
   &&  \left\{
    \int \td k_{1,2,3 \perp}
    T(k_1)T(k_2)T(k_3)
    e^{ix_0\cdot(\sum k_\perp)}
    \right\}
,\end{eqnarray}
which we can write in terms of the D-brane tachyon field \eqref{newtach} as
\begin{eqnarray}
S^{(3)}&=&-\frac13 
     \frac{2(2\pi)^{p+1}}{(4\log 2)^{3/2}\sqrt{\Pcm{K}
        (2\pi)^{p-25}
        \big\langle\Psi_g\big|\Pcm{Q}\Psi_g\big\rangle}} \nn\\
    && \int \td k_{1,2,3 \parallel}
      T_{\Dp}(k_{1\parallel})
      T_{\Dp}(k_{2\parallel})
      T_{\Dp}(k_{3\parallel})
     \delta\left(\sum k_\parallel\right) 
.\end{eqnarray}
The three-tachyon coupling can be read off;
\begin{equation}
\label{gt}
g_T=\frac2{(4\log 2)^{3/2}\sqrt{\Pcm{K} (2\pi)^{p-25}
    \big\langle\Psi_g\big|\Pcm{Q}\Psi_g\big\rangle}}
.\end{equation}
Following \cite{Sen2}, the D$p$-brane tension is given by
\begin{equation}
\Pcm{T}=\frac1{2\pi^2g_T^2}
,\end{equation}
so that using \er{gt} we can recover the standard formula for the ratio of D-brane tensions,
\begin{equation}
\label{tensions}
\frac{\Pcm{T}_{p+1}}{\Pcm{T}_p}=\frac1{2\pi\sqrt{\alpha'}}
,\end{equation}
where we have restored $\alpha'$.
 

\newcommand{\kn}{k_\sharp}
\newcommand{\kb}{k_\flat}

\section{Background $B$-field}
\label{Bfield}

Here we investigate the effect of a constant background $B$-field, and compute
the tension as in the previous section.
The worldsheet CFT is given by the action~\cite{SW}
\begin{equation}
\label{Baction}
S=\frac12\int \td^2z \big(g_{\mu\nu}\partial_\alpha X^\mu\partial_\beta X^\nu h^{\alpha\beta}
    -B_{ij}\partial_\alpha X^i\partial_\beta X^j \epsilon^{\alpha\beta}
    \big)
,\end{equation}
where $\mu$ and $\nu$ run from $0$ to $25$, while for simplicity $i$ and $j$ will
take only the values $24$ and $25$. That is, the $B$-field has non-zero components only
in these two directions.
This action leads us to the usual Neumann boundary conditions for the first $24$ directions, while for the last two, we have
\begin{equation}
\label{ncbc} 
\big(\delta_{ij}\partial_n + B_{ij}\partial_t \big)X^j(z)\bigg|_{\partial\Sigma}=0
.\end{equation}
In the following, we concentrate on these two directions and often omit explicit indices.
The Green function on the unit disc is given by
\begin{equation}
\label{discGrn}
\big\langle X(z)X(w) \big\rangle_{\textup{Disc}} = -\log |z-w|
   -\frac12\left(\frac{1+B}{1-B}\right)\log (z-\frac1{\bar{w}})
   -\frac12\left(\frac{1-B}{1+B}\right)\log ({\bar{z}}-\frac1w)
.\end{equation}
This can be written \cite{SW} in terms of the open-string metric $G$ and the 
non-commutativity parameter $\theta$. These are related to the $B$-field by 
\begin{equation}
\label{GB}
\theta=-\frac1{1+B} B \frac1{1-B} \qquad \textup{and} \qquad G=\frac1{1+B}1\frac1{1-B}
.\end{equation}
On the boundary, $|z|=1$, the correlator \eqref{discGrn} becomes
\begin{equation}
\big\langle X^i(e^{i\phi})X^j(e^{i\phi'}) \big\rangle_{\partial\textup{D}} = 
   -G^{ij}\log \left(2\sin\frac{\phi-\phi'}2\right)^2 + \frac{i}2 \theta^{ij} \epsilon(\phi-\phi')
.\end{equation}


\subsection{$(B,g)$-Parametrisation}

In section \ref{VSFTP}, to define the D-brane state $|\Dp\rangle$ in \er{dps} and the 
tachyon perturbation $|\chi(k)\rangle$ in \er{chi}, we inserted operators $\sigma^\pm$ to 
change the boundary condition to Dirichlet on part of the cylinder $C_n$. We can accommodate 
the new boundary condition \eqref{ncbc} by instead inserting new operators $\beta^\pm$ at the
same positions. Thus on the cylinder $C_n$, we have effectively a Neumann boundary condition
on the region $-\pi/4 < z < \pi/4$, and \er{ncbc} will apply to the rest of the boundary.
We will again absorb the various singular factors due to coincident $\beta^+$ and $\beta^-$ 
operators into the states.

Analogous to \er{dps} we define the sliver state in the presence of a $B$-field as
\begin{equation}
\label{xiB}
\big\langle \Xi_B \big| \phi \big\rangle = \lim_{n \rightarrow \infty} 
                       \Pcm{N}\big\langle f \circ \phi(0) \big\rangle^{\beta}_{C_n}
,\end{equation}
where now the superscript indicates that we have inserted the operators $\beta^\pm$ in
the $24$- and $25$-directions.

We now define a tachyon perturbation in the presence of the $B$-field as
\begin{equation}
\label{Tb}
 \big|T_B\big\rangle = \int \td k n^{-\kn^2-\kb\frac1{1-B^2}\kb } T_B(k) \big|\chi_B(k)\big\rangle 
,\end{equation}
where $\kn$ refers to the first 24 directions, and $\kb$ contains the last two components of $k$.
$T_B(k)$ is the tachyon field, and we have used the notation 
\begin{equation}
\kb\frac1{1-B^2}\kb \equiv \kb^\mu\left(\frac1{1-B^2}\right)_{\mu\nu}\kb^\nu
.\end{equation}
As in the previous section, we can write down the solution to the linearised
equation of motion for the perturbation $\chi_B(k)$;
\begin{equation}
\label{chiB}
\big\langle \chi_B(k) \big| \phi \big\rangle = \lim_{n \rightarrow \infty} 
                       \Pcm{N} n^{2\kn+2\kb\frac1{1-B^2}\kb}\big\langle f \circ \phi(0) 
                       e^{ik\cdot X}(n\pi/4)\big\rangle^{\beta \perp}_{C_n}
,\end{equation}
where this time we have inserted the operators $\beta^\pm$ to impose the $B$-field boundary condition.
Exactly analogously to the case of the D$p$-brane, this state can be shown to satisfy the 
following,
\begin{eqnarray}
\label{OSBseom}
   \big\langle\chi_B(k)\big|\chi_B(k')\big\rangle &=& 
     \Pcm{K}\sqrt{\det(1+B)}
    \lim_{n\rightarrow\infty} n^{2\kn^2+2\kb\frac1{1-B^2}\kb}
    2^{2\kn^2+2\kb\frac1{1-B^2}\kb} \nn\\&&   
    (2\pi)^{26}4\delta(\kn+\kn')
     \det \frac1{1-B^2} \delta\left((\kb+\kb') \det \frac1{1-B^2}  \right) 
\label{OSBseomm} \\
   \big\langle \Xi_B * \chi_B(k) \big| \chi_B(k') \big\rangle &=&
    \Pcm{K}\sqrt{\det(1+B)}
    \lim_{n\rightarrow\infty} n^{2\kn^2+2\kb\frac1{1-B^2}\kb}
    2^{\kn^2+\kb\frac1{1-B^2}\kb}  \nn\\&&
    (2\pi)^{26}4\delta(\kn+\kn')
     \det \frac1{1-B^2} \delta\left((\kb+\kb') \det \frac1{1-B^2}  \right) 
,\end{eqnarray}
so that off-shell
\begin{equation}
\label{OSBseommm}
  \big\langle\chi_B(k)\big|\chi_B(k')\big\rangle =  2^{\kn^2+\kb\frac1{1-B^2}\kb-1}
     \big\langle \chi_B(k) * \Xi_B + \chi_B(k) * \Xi_B \big| \chi_B(k') \big\rangle 
.\end{equation}
We see that the on-shell condition is now
\begin{equation}
\label{onshBc}
 \kn^2+\kb\frac1{1-B^2}\kb=1 
,\end{equation}
to be compared with \er{kseo}. In the limit of a large $B$-field, $1/(1-B^2)\rightarrow 0$, and we
recover the case of a D23-brane with on-shell tachyon condition $\kn^2=k_{\parallel(23)}^2=1$. In the
$B\rightarrow 0$ limit, $1/(1-B^2)\rightarrow 1$ and a D25-brane obtains, 
with $\kn^2+\kb^2=k_{\parallel(25)}^2=1$.
Substituting \er{OSBseom} into the quadratic part of the action as before, 
and considering the large-$B$ limit we are led to
define a 24-dimensional tachyon field by
\begin{equation}
\label{newtachB}
  \widetilde{T_B}(\kn) \equiv 
  \sqrt{\Pcm{K} (2\pi)^2 \big\langle\Psi_g\big|\Pcm{Q}\Psi_g\big\rangle 4\log2 \sqrt{\det (1+B)}  } 
  \int \td^{25-p}\kb T_B(k)
.\end{equation}
The tension of the resulting D$p$-brane (now taking $\kb$ to represent $25-p$ dimensions rather than two) 
is thus given by
\begin{equation}
\label{Btension}
\frac{\Pcm{T}^B_p}{\Pcm{T}_{25}}
 = { \sqrt{\det(1+B)} }{(2\pi)^{(25-p)}{\alpha'}^{(25-p)/2} } 
.\end{equation}


\subsection{$(G,\theta)$-Parametrisation}

Defining the tachyon, in the presence of the open-string metric $G$, we write
\begin{equation}
 \big|T_{NC}\big\rangle = \int \td k n^{k^2 } T_{NC}(k) \big|\chi_{NC}(k)\big\rangle 
,\end{equation}
where since $G$ is the effective metric, it is understood that $k^2 \equiv k^\mu G_{\mu\nu} k^\nu$.
Looking at \er{GB}, we see that this normalisation is the same as that used in \er{Tb}.

The perturbation is now of the form
\begin{equation}
\label{chiNC}
\big\langle \chi_{NC}(k) \big| \phi \big\rangle = \lim_{n \rightarrow \infty}
                       \Pcm{N}n^{k^2}\big\langle f \circ \phi(0)
                       e^{ik\cdot X}(n\pi/4)\big\rangle^{\beta \perp}_{C_n}
,\end{equation}
and the correlators are found to be
\begin{eqnarray}
\label{ncOSBseom}
   \big\langle\chi_{NC}(k)\big|\chi_{NC}(k')\big\rangle &=&
     \Pcm{K}\sqrt{\det G}
    \lim_{n\rightarrow\infty} n^{k^2}
    2^{k^2} 
    (2\pi)^{26}4\delta(k+k')
\label{ncOSBseomm} \\
   \big\langle \Xi_{NC} * \chi_{NC}(k) \big| \chi_{NC}(k') \big\rangle &=&
    \Pcm{K}\sqrt{\det G}
    \lim_{n\rightarrow\infty} n^{2k^2}
    2^{k^2} 
    (2\pi)^{26}4\delta(k+k')
,\end{eqnarray}
with off-shell equation of motion
\begin{equation}
\label{ncOSBseommm}
  \big\langle\chi_{NC}(k)\big|\chi_{NC}(k')\big\rangle =  2^{k^2-1}
     \big\langle \chi_{NC}(k) * \Xi_{NC} + \chi_{NC}(k) * \Xi_{NC} \big| \chi_{NC}(k') \big\rangle 
.\end{equation}
We now have a `non-commutative' on-shell condition 
\begin{equation}
\label{NConshBc}
 k^\mu G_{\mu\nu} k^\nu =1 
,\end{equation}
where we have written $G$ explicitly.
We substitute \er{ncOSBseom} into the quadratic part of the action and re-define the tachyon field $T_{NC}$;
\begin{equation}
\widetilde{T_{NC}} \equiv 
  \sqrt{\Pcm{K} \big\langle\Psi_g\big|\Pcm{Q}\Psi_g\big\rangle \sqrt{\det G} 4\log 2 } T_{NC}
.\end{equation}
With this, the tachyon action becomes
\begin{eqnarray}
 S&=&S^{(2)}+S^{(3)} \nn\\
  &=&-\frac12(2\pi)^{26}\int \td k (k^2-1) \widetilde{T_{NC}}(k) \widetilde{T_{NC}}(-k) \nn\\
  & &-\frac13\frac{2(2\pi)^{26}}{(4\log 2)^{3/2}  \sqrt{\Pcm{K} \big\langle\Psi_g\big|\Pcm{Q}\Psi_g\big\rangle \sqrt{\det G}}} \nn\\
  & & \qquad \int \td k_{1,2,3}  \widetilde{T_{NC}}(k_1) * \widetilde{T_{NC}}(k_2) * \widetilde{T_{NC}}(k_3)  \delta\left(\sum k\right)
\end{eqnarray}
where $*$ represents the Moyal product.
We identify the cubic tachyon coupling to find the tension of this non-commutative D25-brane
\begin{equation}
\label{NCtension}
\Pcm{T}_{25}^{NC} = \sqrt{\det G} \Pcm{T}_{25}
.\end{equation}
Finally, we identify the ratio
\begin{equation}
\frac{\Pcm{T}^B_{23}}{\Pcm{T}_{25}^{NC}} = (2\pi)^2\alpha'\sqrt{\frac{\det (1+B)}{\det G}}=\frac{(2\pi)^2\alpha'}{(\det G)^{1/4}}
,\end{equation}
as demonstrated in the papers \cite{BMM} and \cite{Okuyama}.

\section{Discussion}
\label{discuss}

In \cite{RSZ4} it was proposed that in Vacuum String Field Theory a D-brane of arbitrary dimension may be represented 
by inserting boundary condition-changing twist operators into the CFT
description of the sliver state in the directions transverse to the D-brane.
We have calculated the ratio of tensions of D-branes using this approach and obtain
\er{tensions},  in agreement with standard string theory. Similar calculations
in VSFT have recently been done in \cite{BMM} and \cite{Okuyama}, using the
algebraic oscillator approach.

Additionally, in section \ref{Bfield} turning on a constant $B$-field in some directions can be represented
in similar fashion, by the insertion of operators $\beta^\pm$ which suitably
modify the boundary condition. In this way, we have shown that a non-commutative D-brane
solution can also be constructed from the sliver state, for generic constant $B$-field.

Of course, for the case of vanishing $B$-field, we recover the D25-brane solution, while
in the large-$B$ limit, the boundary conditions become Dirichlet, and we find agreement
with our tension calculations for the lower-dimensional D-brane in section \ref{VSFTP}.
Although for finite $B$ there is no clear way to re-define the tachyon field, in
the D25-brane case we calculated the tension of a non-commutative D25-brane.

There was some discussion in \cite{HK}, \cite{RSZ6} and \cite{RV} as to whether the sliver state
represents a single or multiple D-brane state. Although the quantities calculated here do not
shed light on this question since they are ratios of tensions, they 
do provide further evidence that we are in fact dealing with a D-brane state,
and that the proposals of \cite{RSZ4} describe the correct way to represent
D-branes in VSFT.


\section*{\small Acknowledgements }
P.M. and Y.Y. are grateful to Simon Fraser University for Graduate Fellowship support.
R.R. would like to thank Simon Fraser University for kind hospitality.
This work was supported in part by an operating grant from the
Natural Sciences and Engineering Research Council of Canada.


\setcounter{equation}{0}
\renewcommand{\theequation}{A-\arabic{equation}}
\section*{Appendix}

Here we show the method of calculation of the correlators used in the main text;
specifically, we make an example of \er{chichi}. 
The other correlators used in the main text may be calculated in a similar way,
and the details of the limiting procedure may be found in \cite{RV}.

Starting with \er{Cappcc}, we map $C_{m+n-2}$ to the unit disc 
using $z \rightarrow e^{\frac{4iz}{m+n-2}}$.
We note that since the zero mode of the expansion for $X(z)$ contributes to the
conformal dimension of the tachyon vertex operator, and this is 
absent in the Dirichlet case, the conformal dimension of $e^{ik\cdot X}$ 
(which is of course normal-ordered) is $k_\parallel^2/2$.
On the disc, then,
\begin{equation}
\label{chch}
\big\langle\chi(k)\big|\chi(k')\big\rangle
=\Pcm{N}^2 \lim_{n,m\rightarrow\infty} n^{2k_\parallel^2}m^{{k'}_\parallel^2}
\left(\frac{4i}{m+n-2}\right)^{k_\parallel^2}
\left(\frac{-4i}{m+n-2}\right)^{{k'}_\parallel^2}
\big\langle
e^{ik\cdot X}(1)
e^{ik'\cdot X}(-1)
\big\rangle^{\sigma\perp}_{\textup{Disc}}
.\end{equation}
The zero mode we mentioned above also produces a $\delta$-function in the correlator,~\cite{Pol} giving
\begin{equation}
\big\langle e^{ik\cdot X}(z)e^{ik'\cdot X}(w)\big\rangle^{\sigma\perp}_{\textup{D}}
 = e^{i k\cdot k' \langle X(z)X(w)\rangle}
 \delta(k_\parallel+k_\parallel')
 e^{ix_0\cdot (k_\perp+k_\perp')}
\end{equation}
for the tachyon propagator. The $X$-propagator is given by
\begin{equation}
\big\langle X(z)X(w) \big\rangle_D = -\log |z-w| \pm \log \left|z-\frac1{\bar{w}}\right|
\end{equation}
where for Dirichlet and Neumann directions the $+$ and $-$ sign is used, respectively.
The tachyon correlator on the boundary is thus
\begin{equation}
\label{DiscTwoCorr}
\big\langle e^{ik\cdot X}(e^{i\theta})e^{ik'\cdot X}(e^{i\theta'})\big\rangle^{\sigma\perp}_{\textup{Disc}}
 = \frac{1}{(2\epsilon)^{2h}}\left(2\sin \frac{\theta-\theta'}2\right)^{2k_\parallel\cdot k'_\parallel}
 \delta(k_\parallel+k_\parallel')
 e^{ix_0\cdot (k_\perp+k_\perp')}
\end{equation}
where $x_0$ is the Dirichlet boundary condition. 
Substituting this into \er{chch} we have
(We absorb factors of $\frac{1}{(2\epsilon)^{2h}}$ into the state $\chi(k)$.)
\begin{eqnarray}
\big\langle\chi(k)\big|\chi(k')\big\rangle
&=&\Pcm{N}^2 \lim_{n,m\rightarrow\infty} n^{2k_\parallel^2}m^{2{k'}_\parallel^2}
2^{-2k_\parallel^2}
\left(\frac{4i}{m+n-2}\right)^{k_\parallel^2}
\left(\frac{-4i}{m+n-2}\right)^{{k'}_\parallel^2} \nn\\
&& \delta(k_\parallel+k'_\parallel)e^{ix_0\cdot(k_\perp+k'_\perp)}
.\end{eqnarray}
Taking the $m\rightarrow\infty$ limit, we end up with the result
\begin{equation}
\label{Achichi}
\big\langle\chi(k)\big|\chi(k')\big\rangle
 =\Pcm{K}
 \lim_{n\rightarrow\infty} n^{2k_\parallel^2}
 2^{2k_\parallel^2}
 (2\pi)^{26}\delta(k_\parallel+k_\parallel')e^{ix_0\cdot (k_\perp+k_\perp')}
.\end{equation}



\begin{thebibliography}{99}

\bibitem{Witten1} Edward Witten, \emph{Non-Commutative Geometry and String Field Theory}, Nucl. Phys.~\textbf{B268} (1986) 253-294.
\bibitem{GJ1} David J.~Gross and Antal Jevicki, \emph{Operator Formulation of Interacting String Field Theory (I)}, Nucl.~Phys~\textbf{B283} (1987) 1.
\bibitem{GJ2} David J.~Gross and Antal Jevicki, \emph{Operator Formulation of Interacting String Field Theory (II)}, Nucl.~Phys~\textbf{B287} (1987) 225.
\bibitem{KP} V.~Alan Kosteleck\'y and Robertus Potting, \emph{Analytical construction of a nonperturbative vacuum for the open bosonic string}, Phys.~Rev.~\textbf{D63} (2001) 046007.
\bibitem{GT1} David J.~Gross and Washington Taylor, \emph{Split string Field Theory I}, JHEP \textbf{0108} (2001) 009.
\bibitem{GT2} David J.~Gross and Washington Taylor, \emph{Split string Field Theory II}, JHEP \textbf{0108} (2001) 010.

\bibitem{Sen} Ashoke Sen, \emph{Descent Relations among Bosonic D-branes}, Int.~J.~Mod.~Phys.~\textbf{A14} (1999) 4061.
\bibitem{Sen2} Ashoke Sen, \emph{Universality of the Tachyon Potential},  JHEP~\textbf{9912}~(1999)~027. 

\bibitem{RZ} Leonardo Rastelli and Barton Zwiebach, \emph{Tachyon Potentials, Star Products and Universality}, JHEP \textbf{0109} (2001) 038.

\bibitem{RSZ1} Leonardo Rastelli, Ashoke Sen and Barton Zwiebach, \emph{String Field Theory Around the Tachyon Vacuum}, \texttt{hep-th/0012251}.
\bibitem{RSZ2} Leonardo Rastelli, Ashoke Sen and Barton Zwiebach, \emph{Classical Solutions in String Field Theory Around the Tachyon Vacuum}, \texttt{hep-th/0102112}.
\bibitem{RSZ3} Leonardo Rastelli, Ashoke Sen and Barton Zwiebach, \emph{Half-Strings, Projectors, and Multiple D-branes in Vacuum String Field Theory}, JHEP \textbf{0111} (2001) 035.
\bibitem{RSZ4} Leonardo Rastelli, Ashoke Sen and Barton Zwiebach, \emph{Boundary CFT Construction of D-branes in Vacuum String Field Theory}, JHEP \textbf{0111} (2001) 045.
\bibitem{RSZ5} Leonardo Rastelli, Ashoke Sen and Barton Zwiebach, \emph{Vacuum String Field Theory}, \texttt{hep-th/0106010}.
\bibitem{RSZ6} Leonardo Rastelli, Ashoke Sen and Barton Zwiebach, \emph{A Note on a Proposal for the Tachyon State in Vacuum String Field Theory}, \texttt{hep-th/0111153}.

\bibitem{BMM} L.~Bonora, D.~Mamone and M.~Salizzoni, \emph{B field and squeezed states in Vacuum String Field Theory}, \texttt{hep-th/0201060}.
\bibitem{Okuyama} Kazumi Okuyama, \emph{Ratio of Tensions from Vacuum String Field Theory}, \texttt{hep-th/0201136}.
\bibitem{RV} Radoslav Rashkov and K.~Sankaran Viswanathan, \emph{A Note on the Tachyon State in Vacuum String Field Theory}, \texttt{hep-th/0112202}.
\bibitem{Muk} Partha Mukhopadhyay, \emph{Oscillator Representation of the BCFT Construction of D-branes in Vacuum String Field Theory}, \texttt{hep-th/0110136}.
\bibitem{HK} Hiroyuki Hata and Teruhiko Kawano, \emph{Open String States around a Classical Solution in Vacuum String Field Theory}, \texttt{hep-th/0108150}.
\bibitem{SW} Nathan Seiberg and Edward Witten, \emph{String Theory and Noncommutative Geometry}, JHEP \textbf{9909} (1999) 032. 

\bibitem{Pol} Joseph Polchinski, \emph{String Theory}, Cambridge University Press 1998.

\end{thebibliography}
\end{document}